\documentclass{appolb}
\usepackage[utf8]{inputenc}
\usepackage{epsf}
\usepackage{latexsym,amssymb,euscript}
\usepackage[dvips]{graphicx}
\usepackage[numbers,sort&compress]{natbib}
\usepackage{amsmath}
\usepackage{nicefrac}
\usepackage{slashed}
\usepackage{booktabs}
\usepackage{hyperref}
\usepackage{braket}
\usepackage{chngcntr}
\usepackage{bbold}
\usepackage{graphics}
\usepackage{graphicx}
\usepackage{mciteplus}
\usepackage{caption}
\usepackage{subcaption}
\usepackage{pdfpages}
\usepackage[titletoc]{appendix}
\graphicspath{{./figures/}}
\hypersetup{
 linktocpage = true,
 urlcolor = black,
 colorlinks = true,
 linkcolor = urlblue,
 anchorcolor = urlblue,
 citecolor = urlblue,
 pdfstartview = {XYZ null null 1.25} 
           }
\usepackage{pstricks}
\usepackage{color}
\usepackage{xcolor}
\definecolor{urlblue}{rgb}{0.2,0.4,0.7}
\definecolor{citegreen}{rgb}{0,0.4,0.2}
\definecolor{linkred}{rgb}{0.9,0.2,0.1}
\usepackage{float}
\definecolor{orcidlogocol}{HTML}{A6CE39}
\usepackage{fancyhdr}

\newcommand{\DY}{\Delta Y}

\newcommand{\tref}[1]{~\ref{#1}}

\newcommand{\tarr}{
\begin{array}}
\newcommand{\earr}{\end{array}}

\begin{document}
\title{High-energy signals from heavy-flavor physics%
\thanks{Presented at ``Diffraction and Low-$x$ 2022'', Corigliano Calabro (Italy), September 24-30, 2022.}%
}
\author{
{A.D. Bolognino, M.M.A. Mohammed, A. Papa
\address{Universit\`a della Calabria \& INFN-Cosenza, I-97036 Rende, Cosenza, Italy}
}
\\[3mm]
F.G. Celiberto
\address{ECT*/FBK \& INFN-TIPFA, I-38123 Povo, Trento, Italy}
\\[3mm]
{M. Fucilla\footnote{Speaker.}
\address{Universit\`a della Calabria \& INFN-Cosenza, I-97036 Rende, Cosenza, Italy
\\
Universit\'e Paris-Saclay, CNRS/IN2P3, IJCLab, 91405, Orsay, France}
}
\\[3mm]
D.Yu. Ivanov
\address{Sobolev Institute of Mathematics, 630090 Novosibirsk, Russia}
}
\maketitle
\begin{abstract}
Working in the hybrid high-energy/collinear factorization, where the next-to-leading resummation of energy logarithms is combined with collinear parton densities and fragmentation functions, we study observables sensitive to high-energy dynamics in the context of heavy-flavor physics.
\end{abstract}

\section{Introduction}
\label{sec:intro}

High-energy emissions of heavy-quark flavored objects in hadronic reactions are widely considered as gold-plated probes for the dynamics of fundamental interactions.
Heavy quarks can couple with beyond-the-Standard-Model (BSM) particles. This supports our search for signals of New Physics.
Yet they represent a key ingredient for precision analyses of strong interactions~\cite{Catani:2020kkl,Ball:2022qks,Maltoni:2022bpy,Maciula:2022lzk}, the charm and bottom masses being confined in the perturbative Quantum Chromodynamics (QCD) domain.
In this work we consider the production of a semi-hard heavy-flavored jet emitted in association with a light-flavored one at the LHC~\cite{Bolognino:2021mrc}, as a novel channel for the manifestation of high-energy dynamics.
Then, we provide predictions for the semi-inclusive detection of an ultraforward $D^{* \pm}$ meson at the planned Forward Physics Facility (FPF)~\cite{Anchordoqui:2021ghd,Feng:2022inv} plus a Higgs boson at ATLAS~\cite{Celiberto:2022zdg}.
Our analysis extends the program of high-energy QCD studies (for recent applications, see~\cite{Ducloue:2013hia,Ducloue:2013bva,Caporale:2014gpa,Celiberto:2015yba,Celiberto:2015mpa,Celiberto:2016ygs,Celiberto:2016vva,Caporale:2018qnm,Celiberto:2022gji,Celiberto:2016hae,Celiberto:2016zgb,Celiberto:2017ptm,Celiberto:2017uae,Celiberto:2017ydk,Celiberto:2017ius,Bolognino:2018oth,Bolognino:2019cac,Bolognino:2019yqj,Celiberto:2020rxb,Golec-Biernat:2018kem,Celiberto:2017nyx,Bolognino:2019ouc,Bolognino:2019yls,Bolognino:2019ccd,Celiberto:2021dzy,Celiberto:2021fdp,Bolognino:2022wgl,Boussarie:2017oae,Celiberto:2022keu,Celiberto:2022zdg,Celiberto:2022kza,Bolognino:2021hxxaux,Caporale:2015vya,Caporale:2015int,Caporale:2016soq,Caporale:2016vxt,Caporale:2016xku,Celiberto:2016vhn,Caporale:2016djm,Caporale:2016pqe,Chachamis:2016qct,Chachamis:2016lyi,Caporale:2016lnh,Caporale:2016zkc,Chachamis:2017vfa,Caporale:2017jqj,Celiberto:2020tmb,Celiberto:2021fjf,Celiberto:2021tky,Celiberto:2021txb,Celiberto:2021xpm,Hentschinski:2020tbi,Celiberto:2022fgx,Celiberto:2022qbh}) via the hybrid high-energy/collinear factorization~\cite{Colferai:2010wu,Celiberto:2020wpk,Bolognino:2021mrc,Celiberto:2022rfj,Celiberto:2022dyf,Celiberto:2022kxx} (see also~\cite{Deak:2009xt,Blanco:2020akb,vanHameren:2022mtk,Motyka:2014lya,Brzeminski:2016lwh,Celiberto:2018muu}) built by means of the leading and next-to-leading BFKL resummation~\cite{Fadin:1975cb,Balitsky:1978ic} of energy logarithms (LLA and NLA) and enhanced by the inclusion of collinear parton distribution functions (PDFs) and fragmentation functions (FFs).

\begin{figure}[t]
\centering

   \includegraphics[scale=0.39,clip]{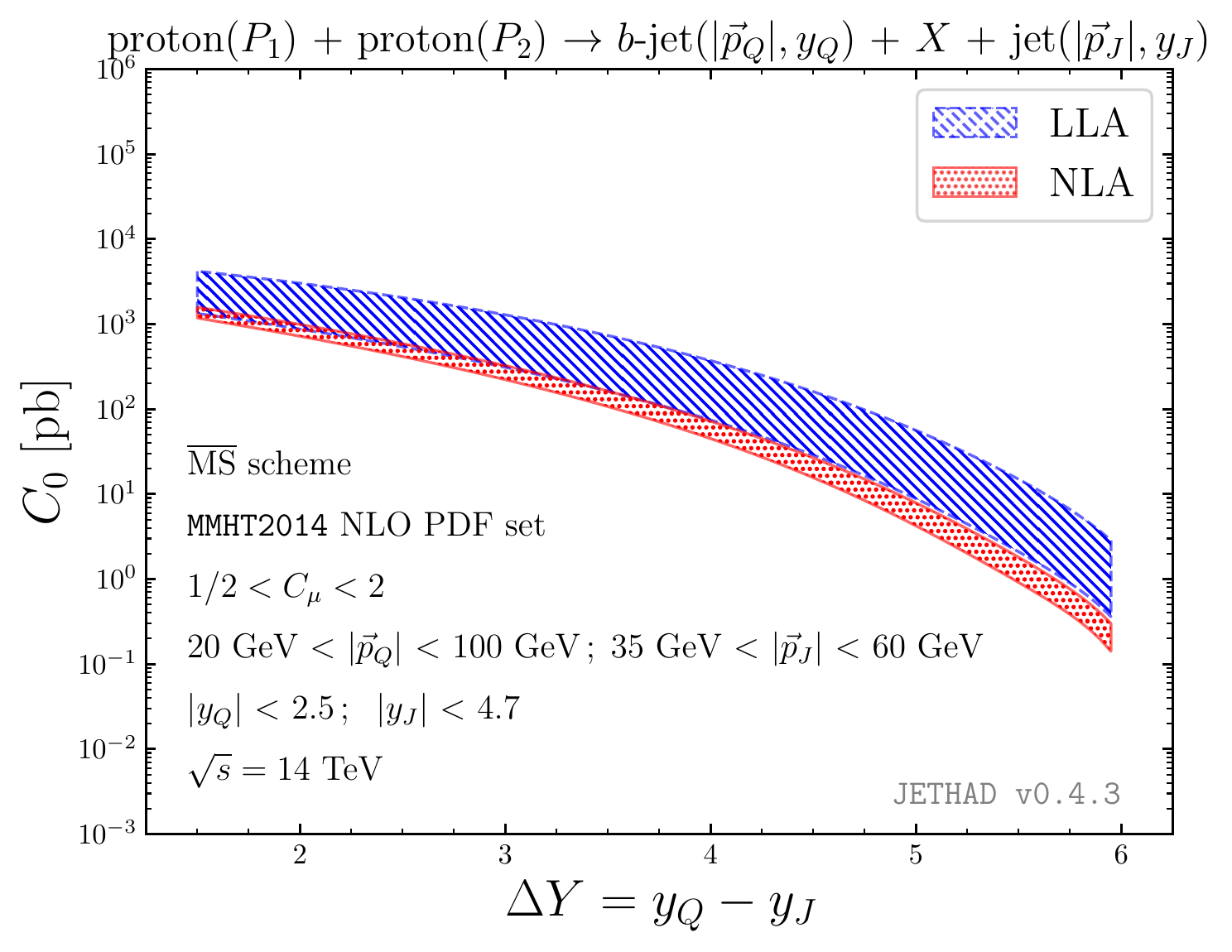}
   \includegraphics[scale=0.39,clip]{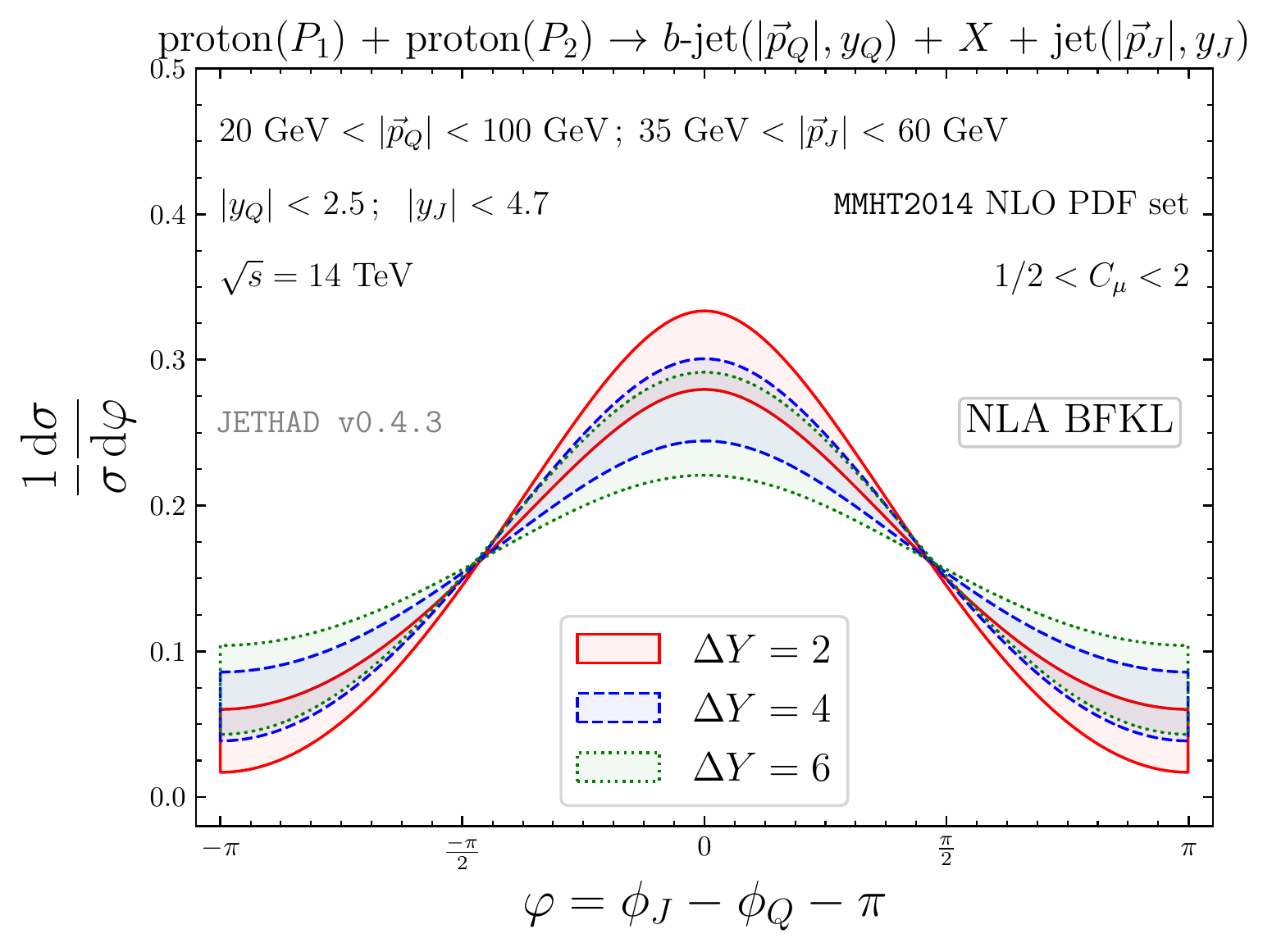}

   \hspace{-0.35cm}
   \includegraphics[scale=0.39,clip]{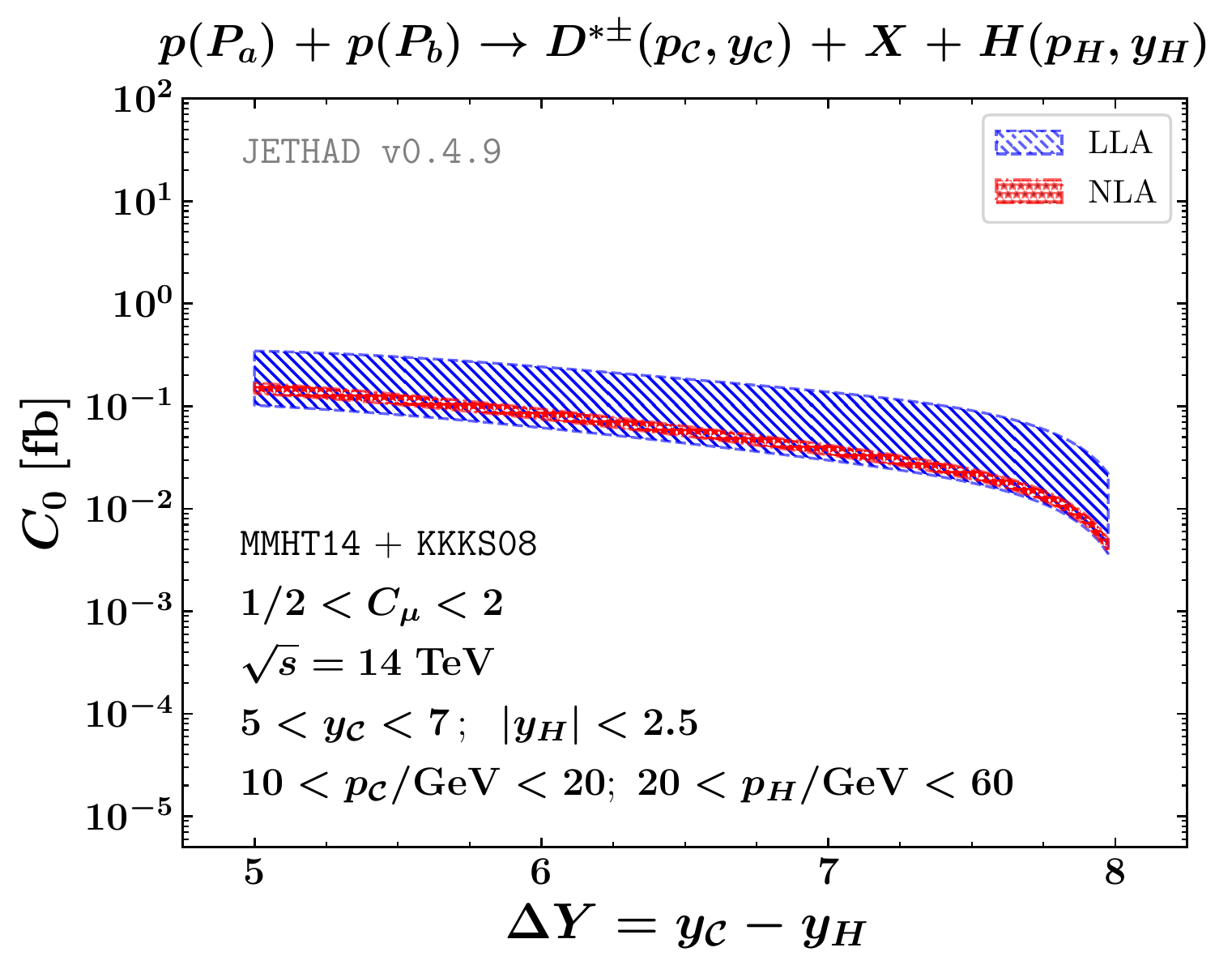}
   \hspace{-0.30cm}
   \includegraphics[scale=0.39,clip]{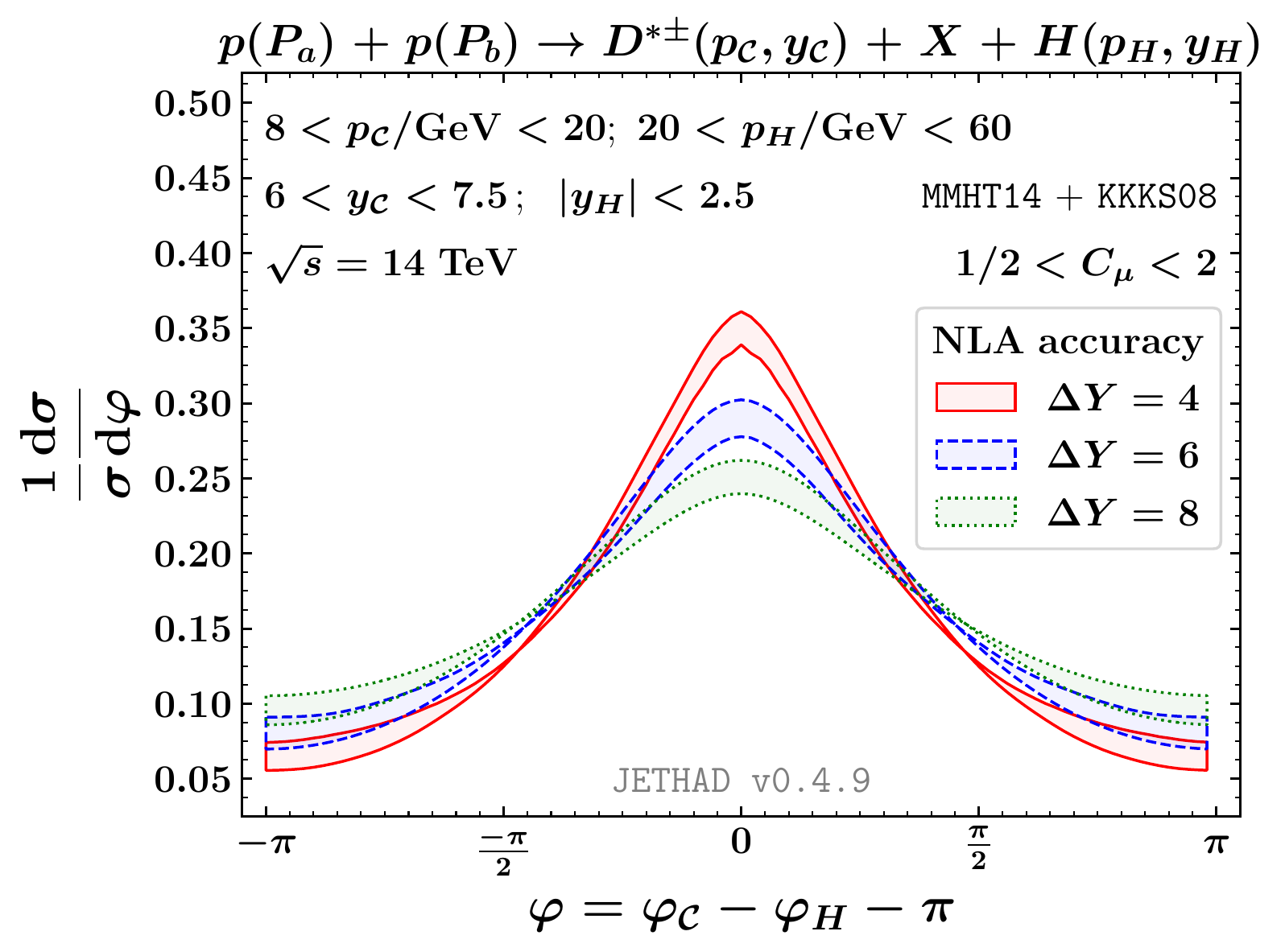}

\caption{Rapidity (left) and azimuthal-angle (right) distributions for the inclusive production of a heavy-light two-jet system at 14 TeV LHC (upper) and of a $D^{* \pm}$ meson plus a Higgs boson at 14 TeV FPF~$+$~ATLAS (lower). Shaded bands refer to the uncertainty coming from scale variation. Text boxes inside panels display final-state kinematic cuts. Figures from~\cite{Bolognino:2021mrc,Celiberto:2022zdg}.}
\label{fig:distributions}
\end{figure}

\section{Phenomenology}
\label{sec:phenomenology}

In left (right) panels of Fig.\tref{fig:distributions} we present NLA rapidity (azimuthal-angle) distributions for the inclusive emission of a heavy-light two-jet system at the LHC (upper panels) and the inclusive $D^{* \pm}$ plus Higgs detection via the FPF~$+$~ATLAS narrow-timing coincidence (lower panels).
NLO proton PDFs are taken from the {\tt NNPDF4.0} set~\cite{NNPDF:2021njg}, while NLO $D^{* \pm}$ FFs are described by the {\tt KKKS08} determination~\cite{Kneesch:2007ey}.
Rapidity distributions exhibit the typical pattern coming from high-energy dynamics.
Here, the increase with energy of partonic cross sections, predicted by BFKL, combines with the dampening effect originating from their convolution with collinear PDFs, thus resulting in a downtrend of distributions as the rapidity distance $\DY$ between the two final-state particles grows.
A clear manifestation of a reached stability under NLA corrections is given by the size of uncertainty bands, which is sensibly lower when passing from LLA to NLA. For some sub-ranges of $\DY$, NLA bands are nested. Moreover, NLA contributions become more and more negative as $\DY$ grows. This is why pure LLA results are always larger.
Azimuthal distributions are shown as functions of $\varphi$, namely the azimuthal-angle distance between the to emitted objects, minus $\pi$.
The peculiar behavior of these observables braces the message that we have entered a kinematic domain where the high-energy resummation works well. All the distributions feature a clear peak at $\varphi = 0$, \emph{i.e.} where the two particles are emitted back-to-back. Then, as $\DY$ grows, the peak height shrinks, whereas the distribution width enlarges. This happens because larger rapidity intervals lead to a stronger decorrelation between the two objects, so that the number of back-to-back events falls off.
As a final remark, we note that all the presented observables present a solid stability under renormalization- and factorization-scale variations at NLA, by which our uncertainty bands are built.
This support the key message that semi-inclusive emissions of heavy-flavored jets as well as bound states lead to a \emph{natural stabilization} (see also~\cite{Celiberto:2021dzy,Celiberto:2021fdp,Celiberto:2022dyf,Celiberto:2022grc}) of the high-energy resummation. This remarkable property represents a core element to lay the groundwork for precision studies of high-energy QCD, and to develop a \emph{multilateral} formalism that combines different resummation mechanisms.

\section{Toward new directions}
\label{sec:conclusions}

The reactions investigated in this work represent a step forward in our ongoing program on heavy-flavored emissions at high energies, started from the analytic calculation of heavy-quark pair impact factors~\cite{Celiberto:2017nyx,Bolognino:2019ouc,Bolognino:2019yls} and pointing toward the analysis of heavy quarkonia~\cite{Boussarie:2017oae,Celiberto:2022dyf,Celiberto:2022keu}. In this context, both two-particle and single-particle detection are relevant. In particular, the second ones will allow us to deepen our understanding of high-energy QCD at new-generation colliders~\cite{AbdulKhalek:2021gbh,Khalek:2022bzd,Chapon:2020heu,Begel:2022kwp,Dawson:2022zbb,Bose:2022obr,Anchordoqui:2021ghd,Feng:2022inv,Hentschinski:2022xnd,Acosta:2022ejc,AlexanderAryshev:2022pkx,Arbuzov:2020cqg,Amoroso:2022eow,Black:2022cth,MuonCollider:2022xlm,Aime:2022flm,MuonCollider:2022ded} via small-$x$ transverse-momentum-dependent gluon densities~\cite{Bolognino:2018rhb,Bolognino:2018mlw,Bolognino:2019bko,Bolognino:2019pba,Celiberto:2019slj,Bolognino:2021niq,Bolognino:2021gjm,Bolognino:2022uty,Celiberto:2022fam,Bolognino:2022ndh,Cisek:2022yjj,Luszczak:2022fkf,Garcia:2019tne,Bacchetta:2020vty,Celiberto:2021zww,Bacchetta:2021oht,Bacchetta:2021lvw,Bacchetta:2021twk,Bacchetta:2022esb,Bacchetta:2022crh,Bacchetta:2022nyv,Celiberto:2022omz,Forte:2015gve,Ball:2017otu,Bonvini:2016wki,Silvetti:2022hyc}.

\bibliographystyle{apsrev}
\bibliography{bibliography}

\begin{thebibliography}{118}
\expandafter\ifx\csname natexlab\endcsname\relax\def\natexlab#1{#1}\fi
\expandafter\ifx\csname bibnamefont\endcsname\relax
  \def\bibnamefont#1{#1}\fi
\expandafter\ifx\csname bibfnamefont\endcsname\relax
  \def\bibfnamefont#1{#1}\fi
\expandafter\ifx\csname citenamefont\endcsname\relax
  \def\citenamefont#1{#1}\fi
\expandafter\ifx\csname url\endcsname\relax
  \def\url#1{\texttt{#1}}\fi
\expandafter\ifx\csname urlprefix\endcsname\relax\def\urlprefix{URL }\fi
\providecommand{\bibinfo}[2]{#2}
\providecommand{\eprint}[2][]{\url{#2}}

\bibitem[{\citenamefont{Catani et~al.}(2021)}]{Catani:2020kkl}
\bibinfo{author}{\bibfnamefont{S.}~\bibnamefont{Catani}} \bibnamefont{et~al.},
  \bibinfo{journal}{JHEP} \textbf{\bibinfo{volume}{03}}, \bibinfo{pages}{029}
  (\bibinfo{year}{2021}), \eprint{2010.11906}.

\bibitem[{\citenamefont{Ball et~al.}(2022{\natexlab{a}})}]{Ball:2022qks}
\bibinfo{author}{\bibfnamefont{R.~D.} \bibnamefont{Ball}} \bibnamefont{et~al.}
  (\bibinfo{collaboration}{NNPDF}), \bibinfo{journal}{Nature}
  \textbf{\bibinfo{volume}{608}}, \bibinfo{pages}{483}
  (\bibinfo{year}{2022}{\natexlab{a}}), \eprint{2208.08372}.

\bibitem[{\citenamefont{Maltoni et~al.}(2022)}]{Maltoni:2022bpy}
\bibinfo{author}{\bibnamefont{Maltoni}} \bibnamefont{et~al.},
  \bibinfo{journal}{JHEP} \textbf{\bibinfo{volume}{10}}, \bibinfo{pages}{027}
  (\bibinfo{year}{2022}), \eprint{2207.10038}.

\bibitem[{\citenamefont{Maciula and Szczurek}(2022)}]{Maciula:2022lzk}
\bibinfo{author}{\bibfnamefont{R.}~\bibnamefont{Maciula}} \bibnamefont{and}
  \bibinfo{author}{\bibfnamefont{A.}~\bibnamefont{Szczurek}}
  (\bibinfo{year}{2022}), \eprint{2210.08890}.

\bibitem[{\citenamefont{Bolognino
  et~al.}(2021{\natexlab{a}})}]{Bolognino:2021mrc}
\bibinfo{author}{\bibfnamefont{A.~D.} \bibnamefont{Bolognino}}
  \bibnamefont{et~al.}, \bibinfo{journal}{Phys. Rev. D}
  \textbf{\bibinfo{volume}{103}}, \bibinfo{pages}{094004}
  (\bibinfo{year}{2021}{\natexlab{a}}), \eprint{2103.07396}.

\bibitem[{\citenamefont{Anchordoqui et~al.}(2022)}]{Anchordoqui:2021ghd}
\bibinfo{author}{\bibfnamefont{L.~A.} \bibnamefont{Anchordoqui}}
  \bibnamefont{et~al.}, \bibinfo{journal}{Phys. Rept.}
  \textbf{\bibinfo{volume}{968}}, \bibinfo{pages}{1} (\bibinfo{year}{2022}),
  \eprint{2109.10905}.

\bibitem[{\citenamefont{Feng et~al.}(2022)}]{Feng:2022inv}
\bibinfo{author}{\bibfnamefont{J.~L.} \bibnamefont{Feng}} \bibnamefont{et~al.}
  (\bibinfo{year}{2022}), \eprint{2203.05090}.

\bibitem[{\citenamefont{Celiberto
  et~al.}(2022{\natexlab{a}})}]{Celiberto:2022zdg}
\bibinfo{author}{\bibfnamefont{F.~G.} \bibnamefont{Celiberto}}
  \bibnamefont{et~al.}, \bibinfo{journal}{Phys. Rev. D}
  \textbf{\bibinfo{volume}{105}}, \bibinfo{pages}{114056}
  (\bibinfo{year}{2022}{\natexlab{a}}), \eprint{2205.13429}.

\bibitem[{\citenamefont{Duclou\'e et~al.}(2013)}]{Ducloue:2013hia}
\bibinfo{author}{\bibfnamefont{B.}~\bibnamefont{Duclou\'e}}
  \bibnamefont{et~al.}, \bibinfo{journal}{JHEP} \textbf{\bibinfo{volume}{05}},
  \bibinfo{pages}{096} (\bibinfo{year}{2013}), \eprint{1302.7012}.

\bibitem[{\citenamefont{Duclou\'e et~al.}(2014)}]{Ducloue:2013bva}
\bibinfo{author}{\bibfnamefont{B.}~\bibnamefont{Duclou\'e}}
  \bibnamefont{et~al.}, \bibinfo{journal}{Phys. Rev. Lett.}
  \textbf{\bibinfo{volume}{112}}, \bibinfo{pages}{082003}
  (\bibinfo{year}{2014}), \eprint{1309.3229}.

\bibitem[{\citenamefont{Caporale et~al.}(2014)}]{Caporale:2014gpa}
\bibinfo{author}{\bibfnamefont{F.}~\bibnamefont{Caporale}}
  \bibnamefont{et~al.}, \bibinfo{journal}{Eur. Phys. J. C}
  \textbf{\bibinfo{volume}{74}}, \bibinfo{pages}{3084} (\bibinfo{year}{2014}),
  \eprint{1407.8431}.

\bibitem[{\citenamefont{Celiberto
  et~al.}(2015{\natexlab{a}})}]{Celiberto:2015yba}
\bibinfo{author}{\bibfnamefont{F.~G.} \bibnamefont{Celiberto}}
  \bibnamefont{et~al.}, \bibinfo{journal}{Eur. Phys. J. C}
  \textbf{\bibinfo{volume}{75}}, \bibinfo{pages}{292}
  (\bibinfo{year}{2015}{\natexlab{a}}), \eprint{1504.08233}.

\bibitem[{\citenamefont{Celiberto
  et~al.}(2015{\natexlab{b}})}]{Celiberto:2015mpa}
\bibinfo{author}{\bibfnamefont{F.~G.} \bibnamefont{Celiberto}}
  \bibnamefont{et~al.}, \bibinfo{journal}{Acta Phys. Polon. Supp.}
  \textbf{\bibinfo{volume}{8}}, \bibinfo{pages}{935}
  (\bibinfo{year}{2015}{\natexlab{b}}), \eprint{1510.01626}.

\bibitem[{\citenamefont{Celiberto
  et~al.}(2016{\natexlab{a}})}]{Celiberto:2016ygs}
\bibinfo{author}{\bibfnamefont{F.~G.} \bibnamefont{Celiberto}}
  \bibnamefont{et~al.}, \bibinfo{journal}{Eur. Phys. J. C}
  \textbf{\bibinfo{volume}{76}}, \bibinfo{pages}{224}
  (\bibinfo{year}{2016}{\natexlab{a}}), \eprint{1601.07847}.

\bibitem[{\citenamefont{Celiberto
  et~al.}(2016{\natexlab{b}})}]{Celiberto:2016vva}
\bibinfo{author}{\bibfnamefont{F.~G.} \bibnamefont{Celiberto}}
  \bibnamefont{et~al.}, \bibinfo{journal}{PoS}
  \textbf{\bibinfo{volume}{DIS2016}}, \bibinfo{pages}{176}
  (\bibinfo{year}{2016}{\natexlab{b}}), \eprint{1606.08892}.

\bibitem[{\citenamefont{Caporale et~al.}(2018)}]{Caporale:2018qnm}
\bibinfo{author}{\bibfnamefont{F.}~\bibnamefont{Caporale}}
  \bibnamefont{et~al.}, \bibinfo{journal}{Nucl. Phys. B}
  \textbf{\bibinfo{volume}{935}}, \bibinfo{pages}{412} (\bibinfo{year}{2018}),
  \eprint{1806.06309}.

\bibitem[{\citenamefont{Celiberto
  et~al.}(2022{\natexlab{b}})}]{Celiberto:2022gji}
\bibinfo{author}{\bibfnamefont{F.~G.} \bibnamefont{Celiberto}}
  \bibnamefont{et~al.}, \bibinfo{journal}{Phys. Rev. D, in press}
  (\bibinfo{year}{2022}{\natexlab{b}}), \eprint{2207.05015}.

\bibitem[{\citenamefont{Celiberto
  et~al.}(2016{\natexlab{c}})}]{Celiberto:2016hae}
\bibinfo{author}{\bibfnamefont{F.~G.} \bibnamefont{Celiberto}}
  \bibnamefont{et~al.}, \bibinfo{journal}{Phys. Rev. D}
  \textbf{\bibinfo{volume}{94}}, \bibinfo{pages}{034013}
  (\bibinfo{year}{2016}{\natexlab{c}}), \eprint{1604.08013}.

\bibitem[{\citenamefont{Celiberto
  et~al.}(2017{\natexlab{a}})}]{Celiberto:2016zgb}
\bibinfo{author}{\bibfnamefont{F.~G.} \bibnamefont{Celiberto}}
  \bibnamefont{et~al.}, \bibinfo{journal}{AIP Conf. Proc.}
  \textbf{\bibinfo{volume}{1819}}, \bibinfo{pages}{060005}
  (\bibinfo{year}{2017}{\natexlab{a}}), \eprint{1611.04811}.

\bibitem[{\citenamefont{Celiberto
  et~al.}(2017{\natexlab{b}})}]{Celiberto:2017ptm}
\bibinfo{author}{\bibfnamefont{F.~G.} \bibnamefont{Celiberto}}
  \bibnamefont{et~al.}, \bibinfo{journal}{Eur. Phys. J. C}
  \textbf{\bibinfo{volume}{77}}, \bibinfo{pages}{382}
  (\bibinfo{year}{2017}{\natexlab{b}}), \eprint{1701.05077}.

\bibitem[{\citenamefont{Celiberto
  et~al.}(2017{\natexlab{c}})}]{Celiberto:2017uae}
\bibinfo{author}{\bibfnamefont{F.~G.} \bibnamefont{Celiberto}}
  \bibnamefont{et~al.} (\bibinfo{year}{2017}{\natexlab{c}}),
  \eprint{1709.01128}.

\bibitem[{\citenamefont{Celiberto
  et~al.}(2017{\natexlab{d}})}]{Celiberto:2017ydk}
\bibinfo{author}{\bibfnamefont{F.~G.} \bibnamefont{Celiberto}}
  \bibnamefont{et~al.} (\bibinfo{year}{2017}{\natexlab{d}}),
  \eprint{1709.04758}.

\bibitem[{\citenamefont{Celiberto}(2017)}]{Celiberto:2017ius}
\bibinfo{author}{\bibfnamefont{F.~G.} \bibnamefont{Celiberto}}, Ph.D. thesis
  (\bibinfo{year}{2017}), \eprint{1707.04315}.

\bibitem[{\citenamefont{Bolognino
  et~al.}(2018{\natexlab{a}})}]{Bolognino:2018oth}
\bibinfo{author}{\bibfnamefont{A.~D.} \bibnamefont{Bolognino}}
  \bibnamefont{et~al.}, \bibinfo{journal}{Eur. Phys. J. C}
  \textbf{\bibinfo{volume}{78}}, \bibinfo{pages}{772}
  (\bibinfo{year}{2018}{\natexlab{a}}), \eprint{1808.05483}.

\bibitem[{\citenamefont{Bolognino
  et~al.}(2019{\natexlab{a}})}]{Bolognino:2019cac}
\bibinfo{author}{\bibfnamefont{A.~D.} \bibnamefont{Bolognino}}
  \bibnamefont{et~al.}, \bibinfo{journal}{PoS}
  \textbf{\bibinfo{volume}{DIS2019}}, \bibinfo{pages}{049}
  (\bibinfo{year}{2019}{\natexlab{a}}), \eprint{1906.11800}.

\bibitem[{\citenamefont{Bolognino
  et~al.}(2019{\natexlab{b}})}]{Bolognino:2019yqj}
\bibinfo{author}{\bibfnamefont{A.~D.} \bibnamefont{Bolognino}}
  \bibnamefont{et~al.}, \bibinfo{journal}{Acta Phys. Polon. Supp.}
  \textbf{\bibinfo{volume}{12}}, \bibinfo{pages}{773}
  (\bibinfo{year}{2019}{\natexlab{b}}), \eprint{1902.04511}.

\bibitem[{\citenamefont{Celiberto et~al.}(2020)}]{Celiberto:2020rxb}
\bibinfo{author}{\bibfnamefont{F.~G.} \bibnamefont{Celiberto}}
  \bibnamefont{et~al.}, \bibinfo{journal}{Phys. Rev. D}
  \textbf{\bibinfo{volume}{102}}, \bibinfo{pages}{094019}
  (\bibinfo{year}{2020}), \eprint{2008.10513}.

\bibitem[{\citenamefont{Golec-Biernat et~al.}(2018)}]{Golec-Biernat:2018kem}
\bibinfo{author}{\bibfnamefont{K.}~\bibnamefont{Golec-Biernat}}
  \bibnamefont{et~al.}, \bibinfo{journal}{JHEP} \textbf{\bibinfo{volume}{12}},
  \bibinfo{pages}{091} (\bibinfo{year}{2018}), \eprint{1811.04361}.

\bibitem[{\citenamefont{Celiberto
  et~al.}(2018{\natexlab{a}})}]{Celiberto:2017nyx}
\bibinfo{author}{\bibfnamefont{F.~G.} \bibnamefont{Celiberto}}
  \bibnamefont{et~al.}, \bibinfo{journal}{Phys. Lett. B}
  \textbf{\bibinfo{volume}{777}}, \bibinfo{pages}{141}
  (\bibinfo{year}{2018}{\natexlab{a}}), \eprint{1709.10032}.

\bibitem[{\citenamefont{Bolognino
  et~al.}(2019{\natexlab{c}})}]{Bolognino:2019ouc}
\bibinfo{author}{\bibfnamefont{A.~D.} \bibnamefont{Bolognino}}
  \bibnamefont{et~al.}, \bibinfo{journal}{PoS}
  \textbf{\bibinfo{volume}{DIS2019}}, \bibinfo{pages}{067}
  (\bibinfo{year}{2019}{\natexlab{c}}), \eprint{1906.05940}.

\bibitem[{\citenamefont{Bolognino
  et~al.}(2019{\natexlab{d}})}]{Bolognino:2019yls}
\bibinfo{author}{\bibfnamefont{A.~D.} \bibnamefont{Bolognino}}
  \bibnamefont{et~al.}, \bibinfo{journal}{Eur. Phys. J. C}
  \textbf{\bibinfo{volume}{79}}, \bibinfo{pages}{939}
  (\bibinfo{year}{2019}{\natexlab{d}}), \eprint{1909.03068}.

\bibitem[{\citenamefont{Bolognino
  et~al.}(2019{\natexlab{e}})}]{Bolognino:2019ccd}
\bibinfo{author}{\bibfnamefont{A.~D.} \bibnamefont{Bolognino}}
  \bibnamefont{et~al.}, \bibinfo{journal}{PoS}
  \textbf{\bibinfo{volume}{DIS2019}}, \bibinfo{pages}{067}
  (\bibinfo{year}{2019}{\natexlab{e}}), \eprint{1906.05940}.

\bibitem[{\citenamefont{Celiberto
  et~al.}(2021{\natexlab{a}})}]{Celiberto:2021dzy}
\bibinfo{author}{\bibfnamefont{F.~G.} \bibnamefont{Celiberto}}
  \bibnamefont{et~al.}, \bibinfo{journal}{Eur. Phys. J. C}
  \textbf{\bibinfo{volume}{81}}, \bibinfo{pages}{780}
  (\bibinfo{year}{2021}{\natexlab{a}}), \eprint{2105.06432}.

\bibitem[{\citenamefont{Celiberto
  et~al.}(2021{\natexlab{b}})}]{Celiberto:2021fdp}
\bibinfo{author}{\bibfnamefont{F.~G.} \bibnamefont{Celiberto}}
  \bibnamefont{et~al.}, \bibinfo{journal}{Phys. Rev. D}
  \textbf{\bibinfo{volume}{104}}, \bibinfo{pages}{114007}
  (\bibinfo{year}{2021}{\natexlab{b}}), \eprint{2109.11875}.

\bibitem[{\citenamefont{Bolognino
  et~al.}(2022{\natexlab{a}})}]{Bolognino:2022wgl}
\bibinfo{author}{\bibfnamefont{A.~D.} \bibnamefont{Bolognino}}
  \bibnamefont{et~al.}, \bibinfo{journal}{PoS}
  \textbf{\bibinfo{volume}{EPS-HEP2021}}, \bibinfo{pages}{389}
  (\bibinfo{year}{2022}{\natexlab{a}}), \eprint{2110.12772}.

\bibitem[{\citenamefont{Boussarie et~al.}(2018)}]{Boussarie:2017oae}
\bibinfo{author}{\bibfnamefont{R.}~\bibnamefont{Boussarie}}
  \bibnamefont{et~al.}, \bibinfo{journal}{Phys. Rev. D}
  \textbf{\bibinfo{volume}{97}}, \bibinfo{pages}{014008}
  (\bibinfo{year}{2018}), \eprint{1709.01380}.

\bibitem[{\citenamefont{Celiberto}(2022{\natexlab{a}})}]{Celiberto:2022keu}
\bibinfo{author}{\bibfnamefont{F.~G.} \bibnamefont{Celiberto}},
  \bibinfo{journal}{Phys. Lett. B} \textbf{\bibinfo{volume}{835}},
  \bibinfo{pages}{137554} (\bibinfo{year}{2022}{\natexlab{a}}),
  \eprint{2206.09413}.

\bibitem[{\citenamefont{Celiberto and Fucilla}(2022)}]{Celiberto:2022kza}
\bibinfo{author}{\bibfnamefont{F.~G.} \bibnamefont{Celiberto}}
  \bibnamefont{and} \bibinfo{author}{\bibfnamefont{M.}~\bibnamefont{Fucilla}},
  \bibinfo{journal}{Zenodo, in press}  (\bibinfo{year}{2022}),
  \eprint{2208.07206}.

\bibitem[{\citenamefont{Bolognino
  et~al.}(2022{\natexlab{b}})}]{Bolognino:2021hxxaux}
\bibinfo{author}{\bibfnamefont{A.~D.} \bibnamefont{Bolognino}}
  \bibnamefont{et~al.}, \bibinfo{journal}{SciPost Phys. Proc.}
  \textbf{\bibinfo{volume}{8}}, \bibinfo{pages}{068}
  (\bibinfo{year}{2022}{\natexlab{b}}), \eprint{2103.07396}.

\bibitem[{\citenamefont{Caporale
  et~al.}(2016{\natexlab{a}})}]{Caporale:2015vya}
\bibinfo{author}{\bibfnamefont{F.}~\bibnamefont{Caporale}}
  \bibnamefont{et~al.}, \bibinfo{journal}{Phys. Rev. Lett.}
  \textbf{\bibinfo{volume}{116}}, \bibinfo{pages}{012001}
  (\bibinfo{year}{2016}{\natexlab{a}}), \eprint{1508.07711}.

\bibitem[{\citenamefont{Caporale
  et~al.}(2016{\natexlab{b}})}]{Caporale:2015int}
\bibinfo{author}{\bibfnamefont{F.}~\bibnamefont{Caporale}}
  \bibnamefont{et~al.}, \bibinfo{journal}{Eur. Phys. J. C}
  \textbf{\bibinfo{volume}{76}}, \bibinfo{pages}{165}
  (\bibinfo{year}{2016}{\natexlab{b}}), \eprint{1512.03364}.

\bibitem[{\citenamefont{Caporale
  et~al.}(2016{\natexlab{c}})}]{Caporale:2016soq}
\bibinfo{author}{\bibfnamefont{F.}~\bibnamefont{Caporale}}
  \bibnamefont{et~al.}, \bibinfo{journal}{Nucl. Phys. B}
  \textbf{\bibinfo{volume}{910}}, \bibinfo{pages}{374}
  (\bibinfo{year}{2016}{\natexlab{c}}), \eprint{1603.07785}.

\bibitem[{\citenamefont{Caporale
  et~al.}(2016{\natexlab{d}})}]{Caporale:2016vxt}
\bibinfo{author}{\bibfnamefont{F.}~\bibnamefont{Caporale}}
  \bibnamefont{et~al.}, \bibinfo{journal}{PoS}
  \textbf{\bibinfo{volume}{DIS2016}}, \bibinfo{pages}{177}
  (\bibinfo{year}{2016}{\natexlab{d}}), \eprint{1610.01880}.

\bibitem[{\citenamefont{Caporale
  et~al.}(2017{\natexlab{a}})}]{Caporale:2016xku}
\bibinfo{author}{\bibfnamefont{F.}~\bibnamefont{Caporale}}
  \bibnamefont{et~al.}, \bibinfo{journal}{Eur. Phys. J. C}
  \textbf{\bibinfo{volume}{77}}, \bibinfo{pages}{5}
  (\bibinfo{year}{2017}{\natexlab{a}}), \eprint{1606.00574}.

\bibitem[{\citenamefont{Celiberto}(2016)}]{Celiberto:2016vhn}
\bibinfo{author}{\bibfnamefont{F.~G.} \bibnamefont{Celiberto}},
  \bibinfo{journal}{Frascati Phys. Ser.} \textbf{\bibinfo{volume}{63}},
  \bibinfo{pages}{43} (\bibinfo{year}{2016}), \eprint{1606.07327}.

\bibitem[{\citenamefont{Caporale
  et~al.}(2017{\natexlab{b}})}]{Caporale:2016djm}
\bibinfo{author}{\bibfnamefont{F.}~\bibnamefont{Caporale}}
  \bibnamefont{et~al.}, \bibinfo{journal}{AIP Conf. Proc.}
  \textbf{\bibinfo{volume}{1819}}, \bibinfo{pages}{060009}
  (\bibinfo{year}{2017}{\natexlab{b}}), \eprint{1611.04813}.

\bibitem[{\citenamefont{Caporale
  et~al.}(2017{\natexlab{c}})}]{Caporale:2016pqe}
\bibinfo{author}{\bibfnamefont{F.}~\bibnamefont{Caporale}}
  \bibnamefont{et~al.}, \bibinfo{journal}{JCEGI} \textbf{\bibinfo{volume}{5}},
  \bibinfo{pages}{47} (\bibinfo{year}{2017}{\natexlab{c}}),
  \eprint{1610.04765}.

\bibitem[{\citenamefont{Chachamis
  et~al.}(2016{\natexlab{a}})}]{Chachamis:2016qct}
\bibinfo{author}{\bibfnamefont{G.}~\bibnamefont{Chachamis}}
  \bibnamefont{et~al.}, \bibinfo{journal}{PoS}
  \textbf{\bibinfo{volume}{DIS2016}}, \bibinfo{pages}{178}
  (\bibinfo{year}{2016}{\natexlab{a}}).

\bibitem[{\citenamefont{Chachamis
  et~al.}(2016{\natexlab{b}})}]{Chachamis:2016lyi}
\bibinfo{author}{\bibfnamefont{G.}~\bibnamefont{Chachamis}}
  \bibnamefont{et~al.} (\bibinfo{year}{2016}{\natexlab{b}}),
  \eprint{1610.01342}.

\bibitem[{\citenamefont{Caporale
  et~al.}(2017{\natexlab{d}})}]{Caporale:2016lnh}
\bibinfo{author}{\bibfnamefont{F.}~\bibnamefont{Caporale}}
  \bibnamefont{et~al.}, \bibinfo{journal}{EPJ Web Conf.}
  \textbf{\bibinfo{volume}{164}}, \bibinfo{pages}{07027}
  (\bibinfo{year}{2017}{\natexlab{d}}), \eprint{1612.02771}.

\bibitem[{\citenamefont{Caporale
  et~al.}(2017{\natexlab{e}})}]{Caporale:2016zkc}
\bibinfo{author}{\bibfnamefont{F.}~\bibnamefont{Caporale}}
  \bibnamefont{et~al.}, \bibinfo{journal}{Phys. Rev. D}
  \textbf{\bibinfo{volume}{95}}, \bibinfo{pages}{074007}
  (\bibinfo{year}{2017}{\natexlab{e}}), \eprint{1612.05428}.

\bibitem[{\citenamefont{Chachamis et~al.}(2018)}]{Chachamis:2017vfa}
\bibinfo{author}{\bibfnamefont{G.}~\bibnamefont{Chachamis}}
  \bibnamefont{et~al.}, \bibinfo{journal}{PoS}
  \textbf{\bibinfo{volume}{DIS2017}}, \bibinfo{pages}{067}
  (\bibinfo{year}{2018}), \eprint{1709.02649}.

\bibitem[{\citenamefont{Caporale
  et~al.}(2017{\natexlab{f}})}]{Caporale:2017jqj}
\bibinfo{author}{\bibfnamefont{F.}~\bibnamefont{Caporale}} \bibnamefont{et~al.}
  (\bibinfo{year}{2017}{\natexlab{f}}), \eprint{1801.00014}.

\bibitem[{\citenamefont{Celiberto
  et~al.}(2021{\natexlab{c}})}]{Celiberto:2020tmb}
\bibinfo{author}{\bibfnamefont{F.~G.} \bibnamefont{Celiberto}}
  \bibnamefont{et~al.}, \bibinfo{journal}{Eur. Phys. J. C}
  \textbf{\bibinfo{volume}{81}}, \bibinfo{pages}{293}
  (\bibinfo{year}{2021}{\natexlab{c}}), \eprint{2008.00501}.

\bibitem[{\citenamefont{Celiberto
  et~al.}(2022{\natexlab{c}})}]{Celiberto:2021fjf}
\bibinfo{author}{\bibfnamefont{F.~G.} \bibnamefont{Celiberto}}
  \bibnamefont{et~al.}, \bibinfo{journal}{SciPost Phys. Proc.}
  \textbf{\bibinfo{volume}{8}}, \bibinfo{pages}{039}
  (\bibinfo{year}{2022}{\natexlab{c}}), \eprint{2107.13037}.

\bibitem[{\citenamefont{Celiberto
  et~al.}(2022{\natexlab{d}})}]{Celiberto:2021tky}
\bibinfo{author}{\bibfnamefont{F.~G.} \bibnamefont{Celiberto}}
  \bibnamefont{et~al.}, \bibinfo{journal}{PoS}
  \textbf{\bibinfo{volume}{EPS-HEP2021}}, \bibinfo{pages}{589}
  (\bibinfo{year}{2022}{\natexlab{d}}), \eprint{2110.09358}.

\bibitem[{\citenamefont{Celiberto
  et~al.}(2022{\natexlab{e}})}]{Celiberto:2021txb}
\bibinfo{author}{\bibfnamefont{F.~G.} \bibnamefont{Celiberto}}
  \bibnamefont{et~al.}, \bibinfo{journal}{PoS}
  \textbf{\bibinfo{volume}{PANIC2021}}, \bibinfo{pages}{352}
  (\bibinfo{year}{2022}{\natexlab{e}}), \eprint{2111.13090}.

\bibitem[{\citenamefont{Celiberto
  et~al.}(2022{\natexlab{f}})}]{Celiberto:2021xpm}
\bibinfo{author}{\bibfnamefont{F.~G.} \bibnamefont{Celiberto}}
  \bibnamefont{et~al.}, \bibinfo{journal}{SciPost Phys. Proc.}
  \textbf{\bibinfo{volume}{10}}, \bibinfo{pages}{002}
  (\bibinfo{year}{2022}{\natexlab{f}}), \eprint{2110.12649}.

\bibitem[{\citenamefont{Hentschinski et~al.}(2021)}]{Hentschinski:2020tbi}
\bibinfo{author}{\bibfnamefont{M.}~\bibnamefont{Hentschinski}}
  \bibnamefont{et~al.}, \bibinfo{journal}{Eur. Phys. J. C}
  \textbf{\bibinfo{volume}{81}}, \bibinfo{pages}{112} (\bibinfo{year}{2021}),
  \eprint{2011.03193}.

\bibitem[{\citenamefont{Celiberto
  et~al.}(2022{\natexlab{g}})}]{Celiberto:2022fgx}
\bibinfo{author}{\bibfnamefont{F.~G.} \bibnamefont{Celiberto}}
  \bibnamefont{et~al.}, \bibinfo{journal}{JHEP} \textbf{\bibinfo{volume}{08}},
  \bibinfo{pages}{092} (\bibinfo{year}{2022}{\natexlab{g}}),
  \eprint{2205.02681}.

\bibitem[{\citenamefont{Celiberto
  et~al.}(2022{\natexlab{h}})\citenamefont{Celiberto, Fucilla, and
  Papa}}]{Celiberto:2022qbh}
\bibinfo{author}{\bibfnamefont{F.~G.} \bibnamefont{Celiberto}},
  \bibinfo{author}{\bibfnamefont{M.}~\bibnamefont{Fucilla}}, \bibnamefont{and}
  \bibinfo{author}{\bibfnamefont{A.}~\bibnamefont{Papa}}
  (\bibinfo{year}{2022}{\natexlab{h}}), \eprint{2209.01372}.

\bibitem[{\citenamefont{Colferai et~al.}(2010)}]{Colferai:2010wu}
\bibinfo{author}{\bibfnamefont{D.}~\bibnamefont{Colferai}}
  \bibnamefont{et~al.}, \bibinfo{journal}{JHEP} \textbf{\bibinfo{volume}{12}},
  \bibinfo{pages}{026} (\bibinfo{year}{2010}), \eprint{1002.1365}.

\bibitem[{\citenamefont{Celiberto}(2021{\natexlab{a}})}]{Celiberto:2020wpk}
\bibinfo{author}{\bibfnamefont{F.~G.} \bibnamefont{Celiberto}},
  \bibinfo{journal}{Eur. Phys. J. C} \textbf{\bibinfo{volume}{81}},
  \bibinfo{pages}{691} (\bibinfo{year}{2021}{\natexlab{a}}),
  \eprint{2008.07378}.

\bibitem[{\citenamefont{Celiberto}(2022{\natexlab{b}})}]{Celiberto:2022rfj}
\bibinfo{author}{\bibfnamefont{F.~G.} \bibnamefont{Celiberto}},
  \bibinfo{journal}{Phys. Rev. D} \textbf{\bibinfo{volume}{105}},
  \bibinfo{pages}{114008} (\bibinfo{year}{2022}{\natexlab{b}}),
  \eprint{2204.06497}.

\bibitem[{\citenamefont{Celiberto
  et~al.}(2022{\natexlab{i}})}]{Celiberto:2022dyf}
\bibinfo{author}{\bibfnamefont{F.~G.} \bibnamefont{Celiberto}}
  \bibnamefont{et~al.}, \bibinfo{journal}{Eur. Phys. J. C}
  \textbf{\bibinfo{volume}{82}}, \bibinfo{pages}{929}
  (\bibinfo{year}{2022}{\natexlab{i}}), \eprint{2202.12227}.

\bibitem[{\citenamefont{Celiberto}(2022{\natexlab{c}})}]{Celiberto:2022kxx}
\bibinfo{author}{\bibfnamefont{F.~G.} \bibnamefont{Celiberto}}
  (\bibinfo{year}{2022}{\natexlab{c}}), \eprint{2208.14577}.

\bibitem[{\citenamefont{Deak et~al.}(2009)}]{Deak:2009xt}
\bibinfo{author}{\bibfnamefont{M.}~\bibnamefont{Deak}} \bibnamefont{et~al.},
  \bibinfo{journal}{JHEP} \textbf{\bibinfo{volume}{09}}, \bibinfo{pages}{121}
  (\bibinfo{year}{2009}), \eprint{0908.0538}.

\bibitem[{\citenamefont{Blanco and others}(2020)\citenamefont{Blanco
  et~al.}}]{Blanco:2020akb}
\bibinfo{author}{\bibfnamefont{E.}~\bibnamefont{Blanco}} \bibnamefont{et~al.},
  \bibinfo{journal}{JHEP} \textbf{\bibinfo{volume}{12}}, \bibinfo{pages}{158}
  (\bibinfo{year}{2020}), \eprint{2008.07916}.

\bibitem[{\citenamefont{van Hameren et~al.}(2022)}]{vanHameren:2022mtk}
\bibinfo{author}{\bibnamefont{van Hameren}} \bibnamefont{et~al.},
  \bibinfo{journal}{JHEP} \textbf{\bibinfo{volume}{11}}, \bibinfo{pages}{103}
  (\bibinfo{year}{2022}), \eprint{2205.09585}.

\bibitem[{\citenamefont{Motyka et~al.}(2015)}]{Motyka:2014lya}
\bibinfo{author}{\bibfnamefont{L.}~\bibnamefont{Motyka}} \bibnamefont{et~al.},
  \bibinfo{journal}{JHEP} \textbf{\bibinfo{volume}{05}}, \bibinfo{pages}{087}
  (\bibinfo{year}{2015}), \eprint{1412.4675}.

\bibitem[{\citenamefont{Brzeminski et~al.}(2017)}]{Brzeminski:2016lwh}
\bibinfo{author}{\bibfnamefont{D.}~\bibnamefont{Brzeminski}}
  \bibnamefont{et~al.}, \bibinfo{journal}{JHEP} \textbf{\bibinfo{volume}{01}},
  \bibinfo{pages}{005} (\bibinfo{year}{2017}), \eprint{1611.04449}.

\bibitem[{\citenamefont{Celiberto
  et~al.}(2018{\natexlab{b}})}]{Celiberto:2018muu}
\bibinfo{author}{\bibfnamefont{F.~G.} \bibnamefont{Celiberto}}
  \bibnamefont{et~al.}, \bibinfo{journal}{Phys. Lett.}
  \textbf{\bibinfo{volume}{B786}}, \bibinfo{pages}{201}
  (\bibinfo{year}{2018}{\natexlab{b}}), \eprint{1808.09511}.

\bibitem[{\citenamefont{Fadin et~al.}(1975)}]{Fadin:1975cb}
\bibinfo{author}{\bibfnamefont{V.~S.} \bibnamefont{Fadin}}
  \bibnamefont{et~al.}, \bibinfo{journal}{Phys. Lett. B}
  \textbf{\bibinfo{volume}{60}}, \bibinfo{pages}{50} (\bibinfo{year}{1975}).

\bibitem[{\citenamefont{Balitsky and Lipatov}(1978)}]{Balitsky:1978ic}
\bibinfo{author}{\bibfnamefont{I.}~\bibnamefont{Balitsky}} \bibnamefont{and}
  \bibinfo{author}{\bibfnamefont{L.}~\bibnamefont{Lipatov}},
  \bibinfo{journal}{Sov.\ J.\ Nucl.\ Phys.} \textbf{\bibinfo{volume}{28}},
  \bibinfo{pages}{822} (\bibinfo{year}{1978}).

\bibitem[{\citenamefont{Ball et~al.}(2022{\natexlab{b}})}]{NNPDF:2021njg}
\bibinfo{author}{\bibfnamefont{R.~D.} \bibnamefont{Ball}} \bibnamefont{et~al.},
  \bibinfo{journal}{Eur. Phys. J. C} \textbf{\bibinfo{volume}{82}},
  \bibinfo{pages}{428} (\bibinfo{year}{2022}{\natexlab{b}}),
  \eprint{2109.02653}.

\bibitem[{\citenamefont{Kneesch et~al.}(2008)}]{Kneesch:2007ey}
\bibinfo{author}{\bibfnamefont{T.}~\bibnamefont{Kneesch}} \bibnamefont{et~al.},
  \bibinfo{journal}{Nucl. Phys. B} \textbf{\bibinfo{volume}{799}},
  \bibinfo{pages}{34} (\bibinfo{year}{2008}), \eprint{0712.0481}.

\bibitem[{\citenamefont{Celiberto}(2022{\natexlab{d}})}]{Celiberto:2022grc}
\bibinfo{author}{\bibfnamefont{F.~G.} \bibnamefont{Celiberto}}
  (\bibinfo{year}{2022}{\natexlab{d}}), \eprint{2211.11780}.

\bibitem[{\citenamefont{A.~Khalek
  et~al.}(2022{\natexlab{a}})}]{AbdulKhalek:2021gbh}
\bibinfo{author}{\bibfnamefont{R.}~\bibnamefont{A.~Khalek}}
  \bibnamefont{et~al.}, \bibinfo{journal}{Nucl. Phys. A}
  \textbf{\bibinfo{volume}{1026}}, \bibinfo{pages}{122447}
  (\bibinfo{year}{2022}{\natexlab{a}}), \eprint{2103.05419}.

\bibitem[{\citenamefont{A.~Khalek et~al.}(2022{\natexlab{b}})}]{Khalek:2022bzd}
\bibinfo{author}{\bibfnamefont{R.}~\bibnamefont{A.~Khalek}}
  \bibnamefont{et~al.} (\bibinfo{year}{2022}{\natexlab{b}}),
  \eprint{2203.13199}.

\bibitem[{\citenamefont{Chapon et~al.}(2022)}]{Chapon:2020heu}
\bibinfo{author}{\bibfnamefont{E.}~\bibnamefont{Chapon}} \bibnamefont{et~al.},
  \bibinfo{journal}{Prog. Part. Nucl. Phys.} \textbf{\bibinfo{volume}{122}},
  \bibinfo{pages}{103906} (\bibinfo{year}{2022}), \eprint{2012.14161}.

\bibitem[{\citenamefont{Begel et~al.}(2022)}]{Begel:2022kwp}
\bibinfo{author}{\bibfnamefont{M.}~\bibnamefont{Begel}} \bibnamefont{et~al.}
  (\bibinfo{year}{2022}), \eprint{2209.14872}.

\bibitem[{\citenamefont{Dawson et~al.}(2022)}]{Dawson:2022zbb}
\bibinfo{author}{\bibfnamefont{S.}~\bibnamefont{Dawson}} \bibnamefont{et~al.}
  (\bibinfo{year}{2022}), \eprint{2209.07510}.

\bibitem[{\citenamefont{Bose et~al.}(2022)}]{Bose:2022obr}
\bibinfo{author}{\bibfnamefont{T.}~\bibnamefont{Bose}} \bibnamefont{et~al.}
  (\bibinfo{year}{2022}), \eprint{2209.13128}.

\bibitem[{\citenamefont{Hentschinski et~al.}(2022)}]{Hentschinski:2022xnd}
\bibinfo{author}{\bibfnamefont{M.}~\bibnamefont{Hentschinski}}
  \bibnamefont{et~al.} (\bibinfo{year}{2022}), \eprint{2203.08129}.

\bibitem[{\citenamefont{Acosta et~al.}(2022)}]{Acosta:2022ejc}
\bibinfo{author}{\bibfnamefont{D.}~\bibnamefont{Acosta}} \bibnamefont{et~al.}
  (\bibinfo{year}{2022}), \eprint{2203.06258}.

\bibitem[{\citenamefont{Adachi et~al.}(2022)}]{AlexanderAryshev:2022pkx}
\bibinfo{author}{\bibfnamefont{I.}~\bibnamefont{Adachi}} \bibnamefont{et~al.}
  (\bibinfo{collaboration}{ILC International Community})
  (\bibinfo{year}{2022}), \eprint{2203.07622}.

\bibitem[{\citenamefont{Arbuzov et~al.}(2021)}]{Arbuzov:2020cqg}
\bibinfo{author}{\bibfnamefont{A.}~\bibnamefont{Arbuzov}} \bibnamefont{et~al.},
  \bibinfo{journal}{Prog. Part. Nucl. Phys.} \textbf{\bibinfo{volume}{119}},
  \bibinfo{pages}{103858} (\bibinfo{year}{2021}), \eprint{2011.15005}.

\bibitem[{\citenamefont{Amoroso et~al.}(2022)}]{Amoroso:2022eow}
\bibinfo{author}{\bibfnamefont{S.}~\bibnamefont{Amoroso}} \bibnamefont{et~al.}
  (\bibinfo{year}{2022}), \eprint{2203.13923}.

\bibitem[{\citenamefont{Black et~al.}(2022)}]{Black:2022cth}
\bibinfo{author}{\bibfnamefont{K.~M.} \bibnamefont{Black}} \bibnamefont{et~al.}
  (\bibinfo{year}{2022}), \eprint{2209.01318}.

\bibitem[{\citenamefont{de~Blas et~al.}(2022)}]{MuonCollider:2022xlm}
\bibinfo{author}{\bibfnamefont{J.}~\bibnamefont{de~Blas}} \bibnamefont{et~al.}
  (\bibinfo{collaboration}{Muon Collider}) (\bibinfo{year}{2022}),
  \eprint{2203.07261}.

\bibitem[{\citenamefont{Aim\`e et~al.}(2022)}]{Aime:2022flm}
\bibinfo{author}{\bibfnamefont{C.}~\bibnamefont{Aim\`e}} \bibnamefont{et~al.}
  (\bibinfo{year}{2022}), \eprint{2203.07256}.

\bibitem[{\citenamefont{Bartosik et~al.}(2022)}]{MuonCollider:2022ded}
\bibinfo{author}{\bibfnamefont{N.}~\bibnamefont{Bartosik}} \bibnamefont{et~al.}
  (\bibinfo{collaboration}{Muon Collider}) (\bibinfo{year}{2022}),
  \eprint{2203.07964}.

\bibitem[{\citenamefont{Bolognino
  et~al.}(2018{\natexlab{b}})}]{Bolognino:2018rhb}
\bibinfo{author}{\bibfnamefont{A.~D.} \bibnamefont{Bolognino}}
  \bibnamefont{et~al.}, \bibinfo{journal}{Eur. Phys. J. C}
  \textbf{\bibinfo{volume}{78}}, \bibinfo{pages}{1023}
  (\bibinfo{year}{2018}{\natexlab{b}}), \eprint{1808.02395}.

\bibitem[{\citenamefont{Bolognino
  et~al.}(2018{\natexlab{c}})}]{Bolognino:2018mlw}
\bibinfo{author}{\bibfnamefont{A.~D.} \bibnamefont{Bolognino}}
  \bibnamefont{et~al.}, \bibinfo{journal}{Frascati Phys. Ser.}
  \textbf{\bibinfo{volume}{67}}, \bibinfo{pages}{76}
  (\bibinfo{year}{2018}{\natexlab{c}}), \eprint{1808.02958}.

\bibitem[{\citenamefont{Bolognino
  et~al.}(2019{\natexlab{f}})}]{Bolognino:2019bko}
\bibinfo{author}{\bibfnamefont{A.~D.} \bibnamefont{Bolognino}}
  \bibnamefont{et~al.}, \bibinfo{journal}{Acta Phys. Polon. Supp.}
  \textbf{\bibinfo{volume}{12}}, \bibinfo{pages}{891}
  (\bibinfo{year}{2019}{\natexlab{f}}), \eprint{1902.04520}.

\bibitem[{\citenamefont{Bolognino et~al.}(2020)}]{Bolognino:2019pba}
\bibinfo{author}{\bibfnamefont{A.~D.} \bibnamefont{Bolognino}}
  \bibnamefont{et~al.}, \bibinfo{journal}{Phys. Rev. D}
  \textbf{\bibinfo{volume}{101}}, \bibinfo{pages}{054041}
  (\bibinfo{year}{2020}), \eprint{1912.06507}.

\bibitem[{\citenamefont{Celiberto}(2019)}]{Celiberto:2019slj}
\bibinfo{author}{\bibfnamefont{F.~G.} \bibnamefont{Celiberto}},
  \bibinfo{journal}{Nuovo Cim.} \textbf{\bibinfo{volume}{C42}},
  \bibinfo{pages}{220} (\bibinfo{year}{2019}), \eprint{1912.11313}.

\bibitem[{\citenamefont{Bolognino
  et~al.}(2021{\natexlab{b}})}]{Bolognino:2021niq}
\bibinfo{author}{\bibfnamefont{A.~D.} \bibnamefont{Bolognino}}
  \bibnamefont{et~al.}, \bibinfo{journal}{Eur. Phys. J. C}
  \textbf{\bibinfo{volume}{81}}, \bibinfo{pages}{846}
  (\bibinfo{year}{2021}{\natexlab{b}}), \eprint{2107.13415}.

\bibitem[{\citenamefont{Bolognino
  et~al.}(2022{\natexlab{c}})}]{Bolognino:2021gjm}
\bibinfo{author}{\bibfnamefont{A.~D.} \bibnamefont{Bolognino}}
  \bibnamefont{et~al.}, \bibinfo{journal}{SciPost Phys. Proc.}
  \textbf{\bibinfo{volume}{8}}, \bibinfo{pages}{089}
  (\bibinfo{year}{2022}{\natexlab{c}}), \eprint{2107.12725}.

\bibitem[{\citenamefont{Bolognino
  et~al.}(2022{\natexlab{d}})}]{Bolognino:2022uty}
\bibinfo{author}{\bibfnamefont{A.~D.} \bibnamefont{Bolognino}}
  \bibnamefont{et~al.} (\bibinfo{year}{2022}{\natexlab{d}}),
  \eprint{2202.02513}.

\bibitem[{\citenamefont{Celiberto}(2022{\natexlab{e}})}]{Celiberto:2022fam}
\bibinfo{author}{\bibfnamefont{F.~G.} \bibnamefont{Celiberto}}
  (\bibinfo{year}{2022}{\natexlab{e}}), \eprint{2202.04207}.

\bibitem[{\citenamefont{Bolognino
  et~al.}(2022{\natexlab{e}})}]{Bolognino:2022ndh}
\bibinfo{author}{\bibfnamefont{A.~D.} \bibnamefont{Bolognino}}
  \bibnamefont{et~al.}, \bibinfo{journal}{Zenodo, in press}
  (\bibinfo{year}{2022}{\natexlab{e}}), \eprint{2207.05726}.

\bibitem[{\citenamefont{Cisek et~al.}(2022)}]{Cisek:2022yjj}
\bibinfo{author}{\bibfnamefont{A.}~\bibnamefont{Cisek}} \bibnamefont{et~al.}
  (\bibinfo{year}{2022}), \eprint{2209.06578}.

\bibitem[{\citenamefont{\L{}uszczak et~al.}(2022)}]{Luszczak:2022fkf}
\bibinfo{author}{\bibfnamefont{A.}~\bibnamefont{\L{}uszczak}}
  \bibnamefont{et~al.}, \bibinfo{journal}{Phys. Lett. B}
  \textbf{\bibinfo{volume}{835}}, \bibinfo{pages}{137582}
  (\bibinfo{year}{2022}), \eprint{2210.02877}.

\bibitem[{\citenamefont{Arroyo~Garcia et~al.}(2019)}]{Garcia:2019tne}
\bibinfo{author}{\bibfnamefont{A.}~\bibnamefont{Arroyo~Garcia}}
  \bibnamefont{et~al.}, \bibinfo{journal}{Phys. Lett. B}
  \textbf{\bibinfo{volume}{795}}, \bibinfo{pages}{569} (\bibinfo{year}{2019}),
  \eprint{1904.04394}.

\bibitem[{\citenamefont{Bacchetta et~al.}(2020)}]{Bacchetta:2020vty}
\bibinfo{author}{\bibfnamefont{A.}~\bibnamefont{Bacchetta}}
  \bibnamefont{et~al.}, \bibinfo{journal}{Eur. Phys. J. C}
  \textbf{\bibinfo{volume}{80}}, \bibinfo{pages}{733} (\bibinfo{year}{2020}),
  \eprint{2005.02288}.

\bibitem[{\citenamefont{Celiberto}(2021{\natexlab{b}})}]{Celiberto:2021zww}
\bibinfo{author}{\bibfnamefont{F.~G.} \bibnamefont{Celiberto}},
  \bibinfo{journal}{Nuovo Cim.} \textbf{\bibinfo{volume}{C44}},
  \bibinfo{pages}{36} (\bibinfo{year}{2021}{\natexlab{b}}),
  \eprint{2101.04630}.

\bibitem[{\citenamefont{Bacchetta
  et~al.}(2022{\natexlab{a}})}]{Bacchetta:2021oht}
\bibinfo{author}{\bibfnamefont{A.}~\bibnamefont{Bacchetta}}
  \bibnamefont{et~al.}, \bibinfo{journal}{SciPost Phys. Proc.}
  \textbf{\bibinfo{volume}{8}}, \bibinfo{pages}{040}
  (\bibinfo{year}{2022}{\natexlab{a}}), \eprint{2107.13446}.

\bibitem[{\citenamefont{Bacchetta
  et~al.}(2022{\natexlab{b}})}]{Bacchetta:2021lvw}
\bibinfo{author}{\bibfnamefont{A.}~\bibnamefont{Bacchetta}}
  \bibnamefont{et~al.}, \bibinfo{journal}{PoS}
  \textbf{\bibinfo{volume}{EPS-HEP2021}}, \bibinfo{pages}{376}
  (\bibinfo{year}{2022}{\natexlab{b}}), \eprint{2111.01686}.

\bibitem[{\citenamefont{Bacchetta
  et~al.}(2022{\natexlab{c}})}]{Bacchetta:2021twk}
\bibinfo{author}{\bibfnamefont{A.}~\bibnamefont{Bacchetta}}
  \bibnamefont{et~al.}, \bibinfo{journal}{PoS}
  \textbf{\bibinfo{volume}{PANIC2021}}, \bibinfo{pages}{378}
  (\bibinfo{year}{2022}{\natexlab{c}}), \eprint{2111.03567}.

\bibitem[{\citenamefont{Bacchetta
  et~al.}(2022{\natexlab{d}})}]{Bacchetta:2022esb}
\bibinfo{author}{\bibfnamefont{A.}~\bibnamefont{Bacchetta}}
  \bibnamefont{et~al.} (\bibinfo{year}{2022}{\natexlab{d}}),
  \eprint{2201.10508}.

\bibitem[{\citenamefont{Bacchetta
  et~al.}(2022{\natexlab{e}})}]{Bacchetta:2022crh}
\bibinfo{author}{\bibfnamefont{A.}~\bibnamefont{Bacchetta}}
  \bibnamefont{et~al.} (\bibinfo{year}{2022}{\natexlab{e}}),
  \eprint{2206.07815}.

\bibitem[{\citenamefont{Bacchetta
  et~al.}(2022{\natexlab{f}})}]{Bacchetta:2022nyv}
\bibinfo{author}{\bibfnamefont{A.}~\bibnamefont{Bacchetta}}
  \bibnamefont{et~al.}, \bibinfo{journal}{Zenodo, in press}
  (\bibinfo{year}{2022}{\natexlab{f}}), \eprint{2208.06252}.

\bibitem[{\citenamefont{Celiberto}(2022{\natexlab{f}})}]{Celiberto:2022omz}
\bibinfo{author}{\bibfnamefont{F.~G.} \bibnamefont{Celiberto}}
  (\bibinfo{year}{2022}{\natexlab{f}}), \eprint{2210.08322}.

\bibitem[{\citenamefont{Forte and Muselli}(2016)}]{Forte:2015gve}
\bibinfo{author}{\bibfnamefont{S.}~\bibnamefont{Forte}} \bibnamefont{and}
  \bibinfo{author}{\bibfnamefont{C.}~\bibnamefont{Muselli}},
  \bibinfo{journal}{JHEP} \textbf{\bibinfo{volume}{03}}, \bibinfo{pages}{122}
  (\bibinfo{year}{2016}), \eprint{1511.05561}.

\bibitem[{\citenamefont{Ball et~al.}(2018)}]{Ball:2017otu}
\bibinfo{author}{\bibfnamefont{R.~D.} \bibnamefont{Ball}} \bibnamefont{et~al.},
  \bibinfo{journal}{Eur. Phys. J. C} \textbf{\bibinfo{volume}{78}},
  \bibinfo{pages}{321} (\bibinfo{year}{2018}), \eprint{1710.05935}.

\bibitem[{\citenamefont{Bonvini et~al.}(2016)}]{Bonvini:2016wki}
\bibinfo{author}{\bibfnamefont{M.}~\bibnamefont{Bonvini}} \bibnamefont{et~al.},
  \bibinfo{journal}{Eur. Phys. J. C} \textbf{\bibinfo{volume}{76}},
  \bibinfo{pages}{597} (\bibinfo{year}{2016}), \eprint{1607.02153}.

\bibitem[{\citenamefont{Silvetti and Bonvini}(2022)}]{Silvetti:2022hyc}
\bibinfo{author}{\bibfnamefont{F.}~\bibnamefont{Silvetti}} \bibnamefont{and}
  \bibinfo{author}{\bibfnamefont{M.}~\bibnamefont{Bonvini}}
  (\bibinfo{year}{2022}), \eprint{2211.10142}.

\end{thebibliography}

\end{document}